\documentclass[cameraready]{Interspeech}
\usepackage{lmodern}
\title{Assessing a Mathematical Model of Syllable Production via CTW Alignment with EMA Data}

\author[affiliation={1}]{Frédéric}{Berthommier}
\address{
$^1$ Univ. Grenoble Alpes, CNRS, Grenoble INP, GIPSA-lab, 38000 Grenoble, France
}
\email{frederic.berthommier@gipsa-lab.grenoble-inp.fr}

\keywords{Articulatory modeling, syllable production, coarticulation, EMA, Maeda parameters, CTW alignment, model validation, articulatory model transparency}

\usepackage{comment}

\usepackage{newtxtext}

\DeclareFontFamilySubstitution{T3}{ntxtlf}{cmr}

\usepackage{tipa}

\let\oldtextipa\textipa
\renewcommand{\textipa}[1]{{\normalfont\oldtextipa{#1}}}

\usepackage{hyperref}
\hypersetup{colorlinks,allcolors=black}
\begin{document}

\maketitle

\begin{abstract}
This study evaluates a mathematical model of syllable production using an EMA dataset from a recently published study, which proved highly compatible with the model's architecture. The data set consists of regularly structured French phrases of equal duration and reduced phonetic and syllabic complexity. A dedicated procedure was used to transform EMA recordings into Maeda parameters. The Model-generated trajectories were then realigned with these transformed data using Canonical Time Warping (CTW). Statistical validation was conducted via a permutation test, comparing alignment quality against a surrogate condition with phonetically incongruent model outputs. The results show significantly better alignment for phonetically congruent pairings, revealing a strong structural correspondence between the model and the articulatory data. These findings provide experimental support for the representational validity of the model.
\end{abstract}

\section{Introduction}
Recent progress in neural speech synthesis has renewed interest in articulatory modeling, raising the question of how different approaches should be evaluated in terms of interpretability and transparency. We propose to distinguish four levels of articulatory model transparency.

At Level~1, end-to-end neural architectures learn articulatory representations directly from data such as EMA~\cite{cho2024}, X-ray~\cite{tabatabaee25b_interspeech}, or real-time MRI~\cite{wu23k_interspeech,lee2026}. These systems often rely on self-supervised speech encoders (e.g., WavLM-type models~\cite{chen2022}) and generate articulatory embeddings with limited interpretability, which can nevertheless be rendered into high-quality speech using neural vocoders such as HiFi-GAN~\cite{kong2020}. This level currently represents the state of the art in perceptual synthesis quality, although the learned parameters remain largely data-driven and are not grounded in articulatory control variables.

Level~2 incorporates explicit articulatory synthesis within a neural framework. Architectures such as Tensor2Tract~\cite{Krug2025-ICASSP} combine neural encoding with a structured vocal-tract synthesizer (e.g., VocalTractLab~\cite{birkholz2013}), where articulatory parameters are constrained by the geometric structure of the model instead of being directly supervised by EMA data. This design ensures physical interpretability alongside competitive acoustic quality. The acoustic--articulatory link is learned statistically but remains grounded in this physical structure. Rather than relying on linguistic annotations, such systems perform large-scale Monte Carlo sampling of plausible articulatory states in VocalTractLab, and a neural motor encoder is then trained in a supervised manner to map WavLM-Large~\cite{chen2022} acoustic embeddings onto articulatory parameters.

At Level~3, gestures are explicitly linked to linguistic content according to principles similar to those of Articulatory Phonology~\cite{browman1992}. The present work relies on a mathematically defined syllabic planning architecture that can implement coarticulation and integrates Maeda articulatory parameters~\cite{maeda1982}. These parameters are not merely geometric descriptors; they approximate biomechanically meaningful control variables~\cite{maeda1994,kroger2022a}. They can be generated directly by the mathematical model, and we also show how to derive them from EMA data through a linear transformation. To compare model-generated trajectories with empirical recordings, Canonical Time Warping (CTW)~\cite{zhou2009} is applied asymmetrically, aligning the model trajectories to the original EMA temporal grid. Although this procedure may introduce substantial local temporal distortions, it preserves the overall syllabic structure. Unlike Levels~1 and~2, this approach does not aim to optimize perceptual synthesis quality, but rather to increase linguistic transparency and provide access to both higher-level and low-dimensional control variables.

Level~4 would correspond to physiological transparency, in which a model generates biomechanical control variables suitable for investigating motor planning and speech control~\cite{kroger2022a}. Such a model aims to reproduce intrinsic temporal delays and trajectory profiles consistent with known physiological constraints. However, the purely mathematical formulation of our model appears to contradict this level of physiological fidelity; nonetheless, it may help address some of the substantial gaps that still exist in this field.

Beyond questions of transparency, this study is motivated by recent experimental evidence reported by Kasper et al. \cite{KASPER2026}, who observed articulatory anticipation emerging at least \SI{200}{\milli\second} prior to segmental onset in regularly structured syllabic sequences. The mathematical model considered here \cite{berthommier2023, berthommier2024, berthommier25_interspeech} was originally designed to generate syllabic organization of the phonetic content through explicit prospective planning of articulatory and formant trajectories. Anticipatory structure is intrinsic to its architecture, and only minor adaptations are required to reproduce the syllabic configurations used in \cite{KASPER2026}.

Structural correspondence between aligned model trajectories and EMA-derived Maeda parameters was evaluated using a permutation-based statistical framework with two conditions, A and B, which allow for syllabic congruence in both but phonetic congruence only in one. The overall methodology therefore combines (i) explicit parametric generation, (ii) asymmetric parametric temporal alignment, and (iii) statistical validation of parametric and formant correspondence. The objective is to assess whether the model captures experimentally observed syllabic organization within a shared articulatory parameter space.

\begin{figure*}[t]
\centering
\includegraphics[scale=0.50]{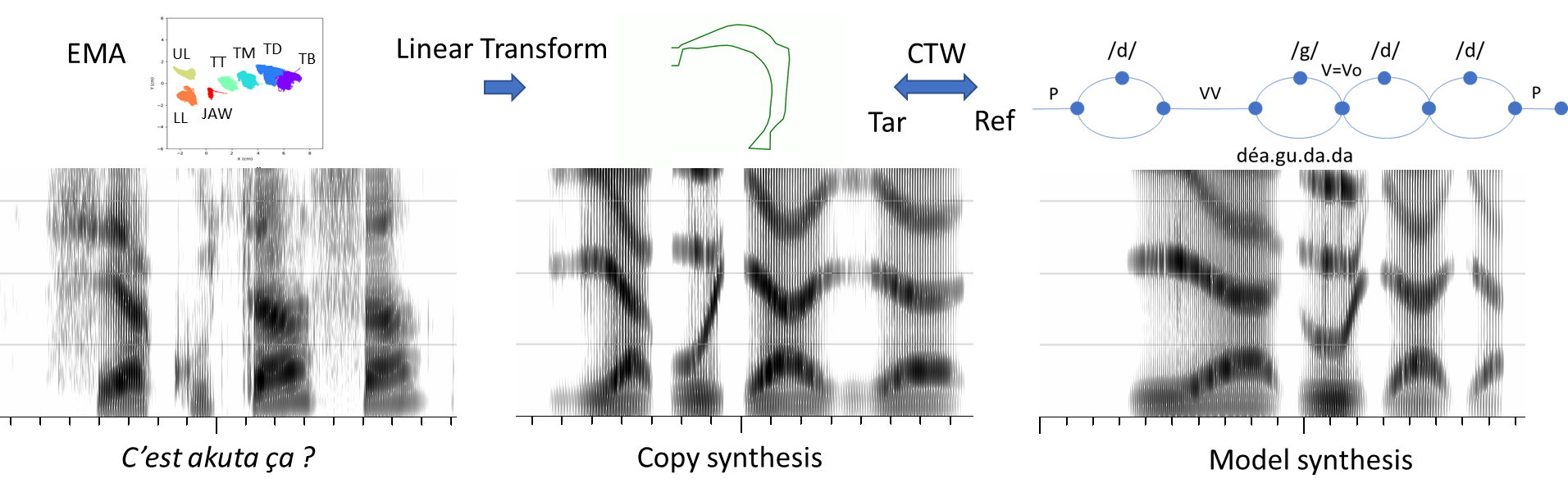}
\caption{Experimental workflow, syllabic graphs, and spectrograms for the example utterance.}
\label{fig:Principle}
\end{figure*}
\section{Methodological details}
The study evaluates whether a deterministic model driven by linguistic content (phoneme and syllabic structures) can produce structural properties that are alignable with EMA recordings of regularly organized phrases. The comparison focuses on the correspondence between grouped modulations of articulatory parameters associated with coarticulation, with the degree of correspondence being determined by the phonetic content.

\subsection{Data and Parameter Space}
The empirical dataset consists of nine repetitions of seven different phrases recorded with EMA by a single speaker. Raw sensor positions were sampled at 500~Hz and transformed into six Maeda-like articulatory parameters using a simple linear mapping procedure. This approach is simpler than in~\cite{toutios2013}, as it does not adjust the geometrical characteristics of the Maeda model based on EMA data. The mapping projects 13 of the 14 EMA coordinates (seven sensors $\times$ two spatial dimensions) onto functionally coherent articulatory degrees of freedom:
\begin{align}
\mathrm{Jaw} &= \mathrm{JAW}_y \\
\mathrm{Body} &= \mathrm{TM}_x + \mathrm{TB}_x + \mathrm{TD}_x \\
\mathrm{Drsm} &= \mathrm{TB}_x + \mathrm{TB}_y + \mathrm{TD}_x + \mathrm{TD}_y \\
\mathrm{Tip} &= \mathrm{TT}_x + \mathrm{TT}_y + \mathrm{TM}_x + \mathrm{TM}_y \\
\mathrm{LipP} &= -(\mathrm{UL}_x + \mathrm{LL}_x) \\
\mathrm{LipH} &= \mathrm{UL}_x + \mathrm{UL}_y + \mathrm{LL}_x - \mathrm{LL}_y.
\end{align}
This linear projection offers a structured form of dimensionality reduction and preserves coordinated articulatory patterns, which appear as pseudoperiodic temporal modulations corresponding to the syllabic rhythm.

After linear transformation, the six-dimensional articulatory trajectories are preprocessed. They are decimated from 500~Hz to 100~Hz (factor of five) to match subsequent analyses; all files contain 1000 frames with approximately 200 silent frames at the end. A zero-phase 8~Hz low-pass filter removes high-frequency noise. Finally, each file is $z$-normalized: for each parameter, the mean is subtracted and the result is divided by the standard deviation.

To enable comparison with the model-generated trajectories, the normalized empirical data, denoted CS, are then rescaled using the mean and standard deviation computed from the corresponding synthetic trajectories produced by the syllabic planning model (described in the following section). Concretely, for each file we compute
\begin{equation}
\mathrm{CS}_{\mathrm{renorm}} = \mathrm{CS}_{\mathrm{norm}} \cdot \sigma_{\mathrm{model}} + \mu_{\mathrm{model}},
\end{equation}
where $\mu_{\mathrm{model}}$ and $\sigma_{\mathrm{model}}$ are the per-parameter mean and standard deviation of the model output for that file. This operation biases the empirical data toward the statistical range of the model and yields a renormalized copy that is later used in dynamic time warping and evaluation. With this simple method, large discrepancies can sometimes appear for the LipH parameter, because a soft rectification is applied in the model at the lip tube level to avoid complete closure. This non‑linearity affects the mapping from LipH to area, which the linear renormalization cannot compensate. The resulting parameters, obtained for each of the 63 files, are used to animate the Maeda model and to recover formant trajectories \cite{BadinFant1984}. Optionally an audio waveform is obtained by retaining the envelope of the original data (see Fig.~\ref{fig:Principle}). This process is referred to as \emph{copy synthesis} in the following.
\vspace{-0.25\baselineskip}
\subsection{Model Generation}
\vspace{-0.25\baselineskip}
\begin{figure*}[t]
\centering
\includegraphics[scale=0.42]{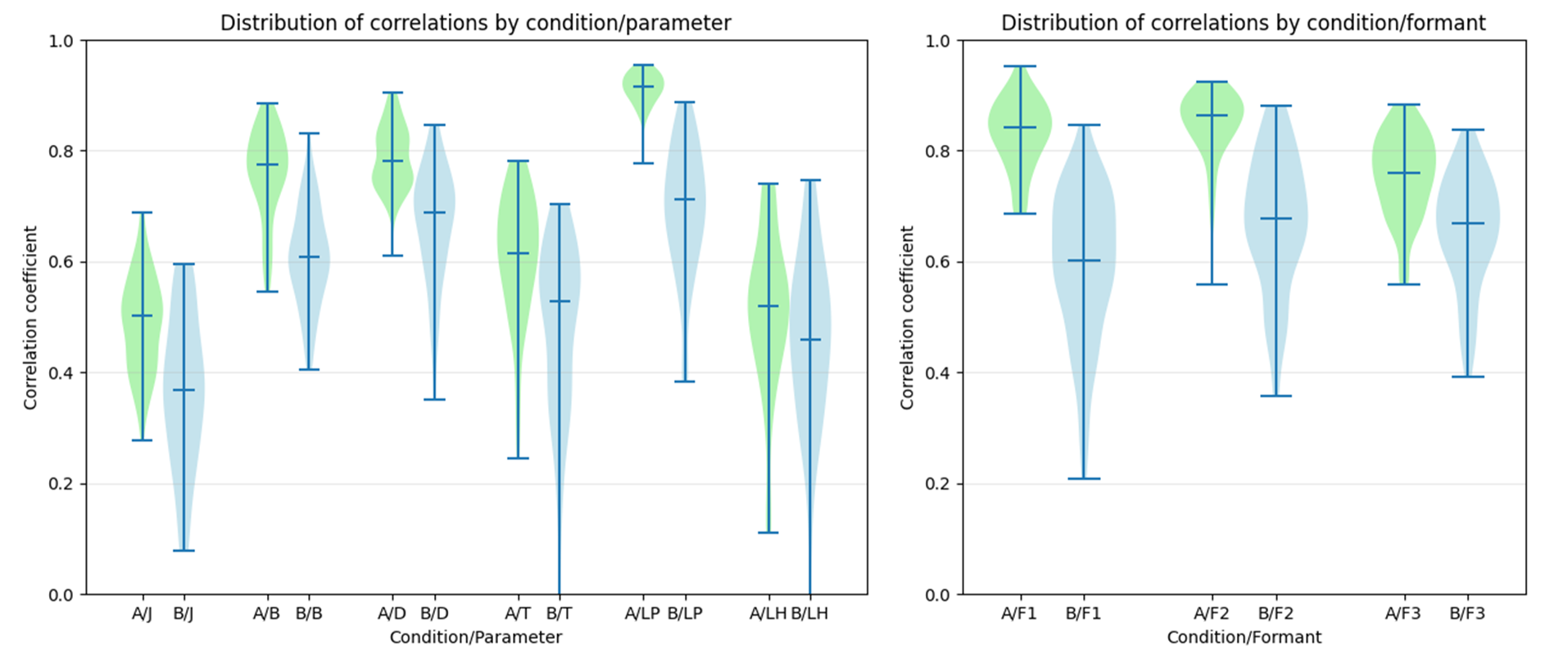}
\caption{Comparison of Conditions A and B across the 63 files. Left: Correlation of Maeda parameters between CS and the CTW-aligned congruent versus incongruent model. Right: Correlation of the synthetic formant trajectories F1–F3.}
\label{fig:Correlations}
\end{figure*}
% \vspace{-12pt}

For each phrase, articulatory trajectories were generated from text inputs specifying the linguistic content, using the syllabic planning model. Syllabification is modeled by directed graphs that encode the information necessary for the selection and coordination of articulators. A syllable graph $G = (V, E)$ consists of nodes representing vowel or consonant targets, and arcs representing transitions. Each node has a location $(\rho,\theta)$ in the complex plane that is phonetically defined and consistent with articulatory features such as degree of constriction and place of articulation.

The phonetic inventory is limited to four vowels /e, a, i, u/ and two consonants /d, g/, which are considered phonetically congruent with /s, t, k/ present in the database. The vowels are positioned as follows:
/e/: $(0.8, 3\pi/2)$, /a/: $(0.8, \pi)$, /i/: $(0.9, 5\pi/3)$, /u/: $(1, \pi/3)$. Consonants are placed slightly outside the unit circle to produce a constriction at an appropriate place of articulation. The consonant /d/ is located at $(1.2, 23\pi/16)$. The consonant /g/ has two allophones: [g\textsubscript{p}] (fronted), associated with the front vowels /e, i/ and placed at $(1.1, 23\pi/12)$, and [g\textsubscript{v}] (retracted), associated with the back vowels /a, u/ and placed at $(1.2, \pi/3)$. Planning is continuous even during silent periods. Syllable graphs are concatenated to form words separated by pauses, ensuring that the articulatory model is permanently animated. Timing is externally controlled by a single parameter $T = 16$ time frames, with an optional different duration for pauses, here $T_p = 16$.

The graph includes two types of arcs: vocalic arcs and consonantal arcs, which are planned simultaneously and then superposed, producing coarticulation without any weighting. For diphthongs (VV) and pauses (P), segments are coordinated across all articulators. Crucially, for other segments, consonants /d, g/ are associated with a specific selection vector $S_c$ that modulates a subset of four articulators (jaw and tongue), while $S_v$ modulates the lip parameters during the vocalic arc (see the supplementary material for details).

A key feature of the graphs is the incorporation of anticipatory vowels, denoted $V_o$, which are not always present in the phonetic string but are essential for smooth transitions. Words are formed by concatenating syllable graphs using the ``.'' operator, which can be implicit (e.g., V.CV). Concatenation is governed by rules for sharing vowel nodes between adjacent syllables, depending on their structure. There are three basic cases, xV.Cx, xC.Cx, and xC.Vx, but only the first one is involved in synthesizing analogs of the data. In xV.Cx, the final vowel of the first syllable serves as the anticipatory vowel $V_o$ for the second syllable, that is, $V_o = V$ for the second syllable.

Each phrase of the database is converted into a series of six analog words compatible with the existing syllable synthesis software. Minor adaptations were introduced to match the regular syllabic patterns. All the words have the same long structure. The first word of Model12, which we will use as the main example (see Fig.~\ref{fig:Principle}), is ``déa.gu.da.da'', resulting from the conversion of the first sequence of ``File 12'' of the database \textit{C'est /\textipa{akuta}/ ça ?} (see the Supplement for the whole set of transcriptions). There are no stationary vowel segments, but the diphthong duration (here ``éa'') is fixed at $2T$. The syllabic graphs appear in Fig.~\ref{fig:Principle} as a series of loops having the consonant node on top and vowel nodes on both sides. The duration of each CV loop is $2T$. Thus, the total duration of one word is $5\times2T=10T$ plus one pause $T_p$ at the beginning. A file contains six such words, and the total duration of its analog is $6\times11\times16=1056$ time frames of \SI{10}{\milli\second}. Consequently, the 7 model sequences (named Model10 to Model16) are slightly longer than the original recorded phrases, which have a duration of \SI{10}{\second}. The model produces six time-varying Maeda parameters used for CTW realignment, as well as corresponding formant trajectories and the audio waveform, as shown in Fig.~\ref{fig:Principle}.

\section{Results}
\vspace{-0.25\baselineskip}
\subsection{Temporal Realignment}
\vspace{-0.25\baselineskip}
To enable frame-wise comparison, CTW~\cite{zhou2009} was applied asymmetrically to warp the model-generated Maeda trajectories onto the temporal grid of the EMA-derived Maeda parameters (1000 samples), leaving these unaltered. For each time index of the EMA-derived Maeda parameter sequence, the aligned model value was obtained by averaging the model frames mapped to that index by the CTW path. The resulting aligned model trajectories were then low-pass filtered at 8 Hz using a zero-phase filter to remove alignment artifacts. Because the last 200 frames of each sequence are silent, the active speech region corresponds to the first 800 frames (indices 0–800). The 63 warping paths obtained from the realignment procedure in conditions~A and~B differ notably in their dispersion over this active region (see Fig.~S1). In condition~A, the pathways remain close to a straight line (after scaling for the different lengths of model and active speech regions), indicating good alignment quality. In contrast, condition~B exhibits larger deviations, reflecting phonetic incongruence. This unique CTW warping path applies simultaneously to all six articulatory parameters, preserving their coordinated temporal structure.

\begin{figure*}[t]\centering\includegraphics[scale=0.43]{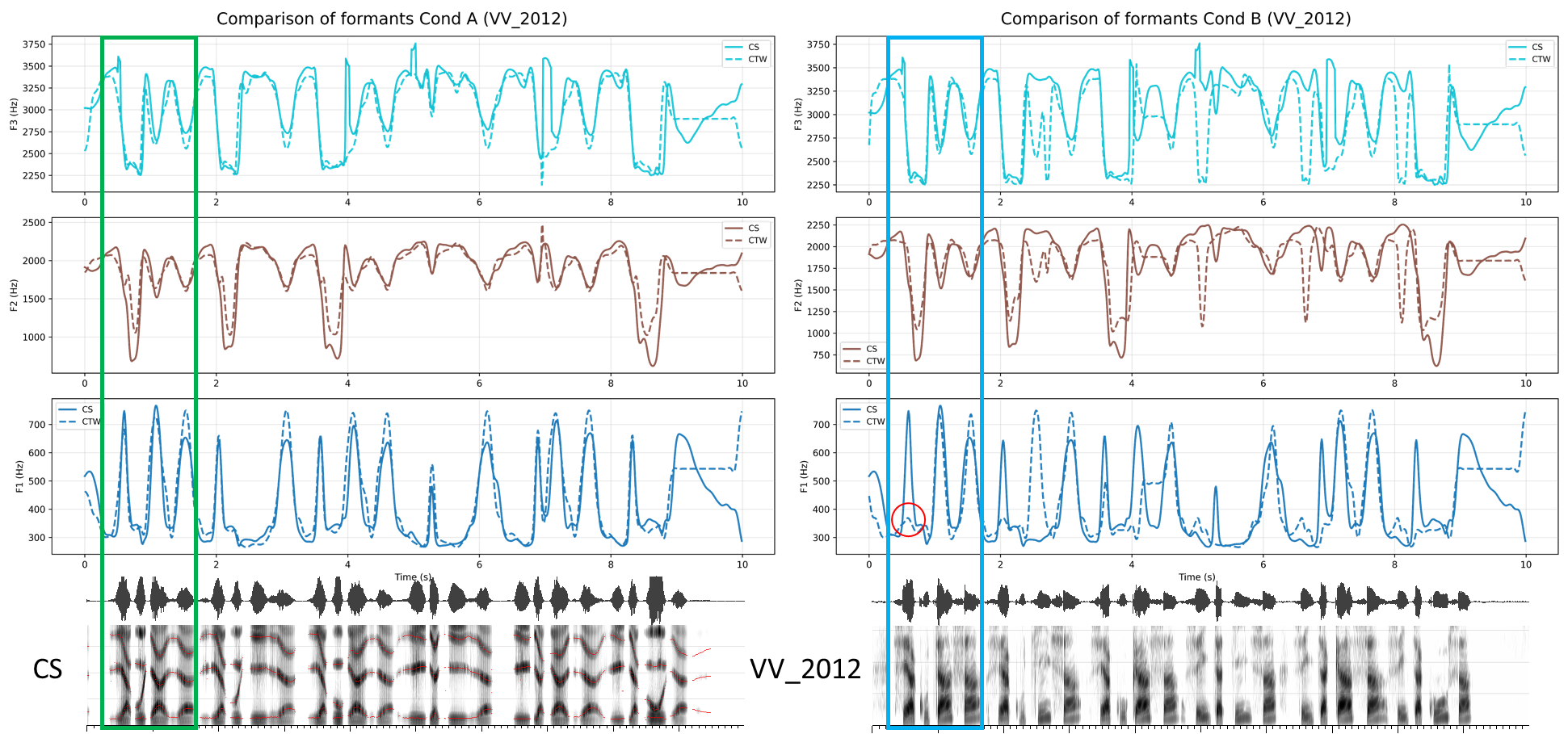}\caption{Comparison between formant trajectories F1-F3 produced by copy synthesis (CS) and those obtained after model realignment, for congruent (Condition~A, left) and incongruent (Condition~B, right) conditions. The CS and speech spectrograms are shown for visual reference.}\label{fig:Formants}\end{figure*}
\vspace{-20pt} 

\vspace{1.5\baselineskip}
\subsection{Correlations between CS and Modeled Trajectories}
\vspace{-0.5\baselineskip}
We conducted a comparative analysis of correlation coefficients obtained between (i) the EMA-derived Maeda parameters (copy synthesis, CS) and (ii) the syllabic model-generated Maeda parameters after CTW alignment, under congruent (Condition~A) and incongruent (Condition~B) pairings. Figure~\ref{fig:Correlations} presents violin plots of the parameter-wise correlations (left) and of the reconstructed formant correlations F1--F3 (right), contrasting both conditions across 63 reference phrases.

Across all six articulatory parameters (Jaw, Body, Dorsum, Tip, LipP, LipH), Condition~A systematically yielded higher mean correlations than Condition~B (Fig.~\ref{fig:Correlations}, left). Paired $t$-tests with Holm correction confirmed that all differences were statistically significant (corrected $p < 0.05$). The largest effects were observed for LipP ($\Delta = 0.2195$, $t = 13.947$), Body ($\Delta = 0.1506$, $t = 14.365$), and Dorsum ($\Delta = 0.1192$, $t = 10.093$), while Jaw ($\Delta = 0.1216$, $t = 7.348$), Tip ($\Delta = 0.1250$, $t = 7.159$), and LipH ($\Delta = 0.0519$, $t = 2.240$) also showed significant improvements in the congruent condition.

A similar pattern was observed at the formant trajectory level following Maeda model resynthesis, using both copy synthesis and aligned model parameters (Fig.~\ref{fig:Correlations}, right). For F1, F2, and F3, Condition~A again produced significantly higher correlations than Condition~B (all Holm-corrected $p < 0.05$). The largest difference was found for F1 ($\Delta = 0.2375$, $t = 13.585$), followed by F2 ($\Delta = 0.1747$, $t = 11.283$) and F3 ($\Delta = 0.0998$, $t = 6.794$). Mean correlations in Condition~A were relatively high (F1: 0.83; F2: 0.85; F3: 0.75), indicating strong spectral reconstruction under congruent alignment.

Correlations in Condition~B remained substantial despite phonetic incongruence, which can be attributed to preserved syllabic structure and to shared lexical material (“C’est” and “ça”) across phrases. This suggests that CTW alignment preserves higher-level syllabic organization, allowing substantial structural matching even when segmental targets differ. 

Finally, although Jaw and LipH exhibited comparatively lower articulatory correlations in Condition A (probably due to the linear estimation as mentioned before), the acoustic reconstruction—particularly for F2—remained robust. This is consistent with the reduced sensitivity of F2 to vertical jaw displacement and lip height compared to tongue body and dorsum configurations. As a control, renormalizing CS parameters in Condition~B with the statistics of the randomly paired incongruent model yielded a similar contrast with Condition~A (see Fig.~S2). Overall, these results demonstrate that phonetic congruence drives significantly better articulatory and acoustic correspondence, whereas the preserved syllabic structure common to both conditions only sustains a baseline level of alignment. 

\subsection{Detailed Analysis of CTW Alignment}
\vspace{-0.5\baselineskip}
We illustrate the behavior of CTW under two alignment conditions (A and B) using the initial segment of the phrase \texttt{VV\_2012} \textit{C'est /\textipa{akuta}/ ça ?} (see Fig.~\ref{fig:Principle}). In condition~A, the model-generated Maeda parameters of Model12—whose initial segment is derived from the syllabic analogue ``déa.gu.da.da''—are aligned onto the target Maeda parameters obtained by copy synthesis (CS). In condition~B, the model-generated Maeda parameters of Model15—beginning with ``déu.da.di.da'', which corresponds to the onset of ``File 15'' \textit{C'est /\textipa{utati}/ ça ?}—are aligned onto the target CS parameters, thereby introducing systematic phonetic differences. These differences are directly reflected in the modulation of the Maeda parameters of Model12 and Model15. Although both conditions share similar syllabic timing, the alignment between modulations of the CS parameters and synthetic modulations will crucially depend on their similarity, which is controlled here by the phonetic content. Although the synthetic modulations do not perfectly reproduce the CS modulations, this test effectively probes their capacity to be aligned with an a priori congruent model. As shown in Fig.~\ref{fig:Formants}, which presents the resulting formant trajectories after alignment, CTW globally compensates for most structural discrepancies in condition~B. The primary mismatch concerns the first formant (F1) of the vowel /a/ in ``déa'', which is not reproduced in condition~B because the corresponding segment is /u/ in ``déu''. This absence is consistent with the lack of lip opening (LipH) and jaw lowering (Jaw) observed in the aligned articulatory parameters (see Fig.~S3). In contrast, the remaining parameters exhibit substantial temporal realignment. Further inspection of Model12 and Model15 (see Fig.~S4 and Fig.~S5) reveals that their articulatory parameters are largely opposite in phase, which explains why Model15 does not match CS. Additionally, CTW resolves structural discrepancies by deleting the syllable /di/ in condition~B through many-to-one temporal mapping. Overall, this example illustrates that CTW effectively probes phonetic incongruencies at the parametric level, demonstrating a high degree of interpretability related to the modelling of coarticulation phenomena.
\vspace{-0.5\baselineskip}
\section{Conclusion}
This experiment demonstrates that articulatory parameters can be generated by a mathematical model and aligned with natural articulatory data in the context of regular and controlled linguistic content. The model goes beyond traditional rule-based synthesis by reproducing temporal modulations observed at the level of articulatory parameters. Furthermore, the experiment shows that CTW is a promising tool for aligning articulatory data by leveraging these modulations and for probing phonetic congruence at the parametric level.

\section{Data Availability}
The data and materials used in this study are openly available. The dataset is available on both Zenodo (\url{https://doi.org/10.5281/zenodo.22961484}) and GitHub (\url{https://github.com/FBerthommier/EMA-to-Maeda}). The supplementary material, also referred to in the text as the Supplement and includes Figures S1--S5, is available on Zenodo (\url{https://doi.org/10.5281/zenodo.22961484}). The code and all resources required for reproducibility are available on GitHub (\url{https://github.com/FBerthommier/EMA-to-Maeda}).

\section{Acknowledgments}
I would like to thank Pascal Perrier, a co-author of Kasper et al., for personally sharing the data with me. 

\bibliographystyle{IEEEtran}
\bibliography{mybib}

\end{document}